\documentclass[11pt]{article}
\usepackage{coling2020}
\usepackage{times}
\usepackage{url}
\usepackage{latexsym}
\usepackage{hyperref}
\usepackage{graphicx}
\usepackage{subcaption}
\usepackage{comment}

\usepackage{floatrow}
\newfloatcommand{capbtabbox}{table}[][\FBwidth]

\title{Social Analysis of Young Basque Speaking Communities in Twitter
\thanks{This article has been accepted for publication in Journal of Multilingual and Multicultural Development, published by Taylor \& Francis.} \thanks{Joseba Fernandez de Landa and Rodrigo Agerri. Social analysis of young Basque-speaking communities in twitter. Journal of Multilingual and Multicultural Development (2021). \url{https://doi.org/10.1080/01434632.2021.1962331}.}
}

\author{Joseba Fernandez de Landa , Rodrigo Agerri \\
  HiTZ Center - Ixa, University of the Basque Country UPV/EHU \\
  {\tt \{joseba.fernandezdelanda,rodrigo.agerri\}@ehu.eus}
  }

\begin{document}
\maketitle
\begin{abstract}
In this paper we take into account both social and linguistic aspects to
perform demographic analysis by processing a large amount of tweets in Basque
language. The study of demographic characteristics and social relationships
are approached by applying machine learning and modern deep-learning Natural Language Processing (NLP)
techniques, combining social sciences with automatic text processing. More
specifically, our main objective is to combine demographic inference and social
analysis in order to detect young Basque Twitter users and to identify the
communities that arise from their relationships or shared content. 
This social and demographic analysis will be entirely based on
the~automatically collected tweets using NLP to convert unstructured textual
information into interpretable knowledge.
\end{abstract}

\section{Introduction}\label{sec:intro}

Basque is a low resourced language, spoken by 28.4\% and understood by 44.8\%
of the population of the Basque Country \cite{inksoz2016sei}. Thanks to its
official status, it is a language with presence in the regional
public administration, education system and in some news media. Thus, in EiTB
(Euskal Irrati Telebista), the Basque public radio and television broadcaster,
it is possible to find radio and television channels in which all content is
entirely broadcasted in Basque. Furthermore, there are other independent media
such as Berria (newspaper), Argia (weekly magazine) and HamaikaTB
(a television channel), in which Basque is the vehicular language. Still, the
presence of Basque in traditional television and news media remains quite low,
particularly when compared with those available for Spanish.

In this context, the increasingly used social networks such as Twitter are of
particular importance for a low resourced language such as Basque. Thus, it is
possible to find a strong and active community of Basque speakers in Twitter
which generates, for a low resourced language, a large amount of textual
content written on Basque. Furthermore, as users create both explicit and
implicit relations and communities, this data is useful to do social research
using methods that may complement those traditionally used in sociology
\cite{baldwin2015shared,nguyen2016computational,rosenthal2017semeval}.
Following this, a promising and relatively new avenue of research in social and
demographic analysis combines the study of social structures created in media
such as Twitter with the automatic analysis of texts via NLP. For example, previous 
work has focused on Twitter to study
the spread of rumours \cite{derczynski2017semeval}, the detection of political
stance \cite{mohammad-etal-2016-semeval} or hate speech
\cite{basile-etal-2019-semeval}.

In this paper we will take into account both
social and linguistic aspects in order to perform demographic analysis by
processing a large amount of tweets in Basque language. The study of
demographic characteristics and social relationships will be approached by
applying machine learning and modern deep-learning NLP techniques, combining social sciences with
automatic text processing.

More specifically, our main objective is to combine demographic inference and
social analysis in order to detect young Basque Twitter
users and to identify the communities that arise from their relationships or shared
content.  By ``Basque Twitter users'' we refer to those that write at least 20\%
of their tweets in Basque.  This social and demographic analysis will be
entirely based on the~automatically collected tweets using NLP to convert
unstructured textual information into interpretable knowledge.

Current work substantially improves and extends the preliminary experimental
work presented in \cite{fernandez2019large}. These improvements have led to
a number of contributions. First, and taking as a starting point
the Heldugazte corpus containing 6M tweets in Basque language
\cite{fernandez2019large}, we devise a whole new methodology to classify users
by life-stage (young/adult). This new method generates a new
dataset, \emph{Heldugazte-Age}, containing 80K tweets semi-automatically annotated at young/adult level. 
Second, we explore the application of modern pre-trained large multilingual and
monolingual models \cite{Devlin19,agerri-etal-2020-give} in order to identify
young and adult users. Third, we perform a qualitative analysis comparing
human performance vs life stage classifiers for classifying Basque users into
the young or adult categories. Four, we use recently developed deep learning
techniques for community detection, achieving better detection and
visualization of the communities, as well as providing information of the
relations among them. We believe that the methodology presented in this paper
might be of interest for other NLP tasks and other types of social and
demographic studies. Finally, we publicly distribute every resource (software
and data) to facilitate further research for low resourced
languages such as
Basque\footnote{\url{https://github.com/ixa-ehu/heldugazte-corpus}}.

The rest of the paper is structured as follows. In the next section we describe
related work in computational sociolinguistics and NLP. In Section
\ref{sec:heldugazte-age} we present our method to build the
\emph{Heldugazte-Age} annotated dataset to train classifiers for life stage
detection. Section \ref{sec:experimental-setup} presents systems used to train
classifiers for life stage detection. These classifiers are evaluated in
Section \ref{sec:life-stage-detection} and applied to perform social network
analysis in Section \ref{sec:relationship-network}. We finish with some
concluding remarks and future work.

\section{Context and Related Work}\label{sec:related}

Social media offers the opportunity to express beliefs,
sentiments or opinions in a variety of formats, including text, image, audio
and video. Social media publications express conscious and/or subconscious
manifestations of our social, emotional and rational condition. 

Previous work in sociolinguistics argues that our writing style can even be a reflection of
demographic characteristics \cite{nguyen2016computational}. Considering the
fact that language is a social phenomenon and thanks to the ever-growing
capacity in the NLP field to collect and process
large-scale amounts of texts, computational sociolinguistics is becoming
increasingly popular. The widespread use of Twitter
has in fact benefited such approaches as it is possible now to mine large
amounts of texts also for less resourced languages.

Twitter is widely used in NLP for tasks
such as mining opinions about specific products or topics
\cite{villena2013tass,rosenthal2017semeval}, detecting political stance
\cite{mohammad-etal-2016-semeval,derczynski2017semeval} and hate speech \cite{basile-etal-2019-semeval} or for basic tasks
such as POS tagging \cite{ritter_named_2011}, named entity recognition
\cite{baldwin2015shared}, normalization~\cite{alegria2015tweetnorm}
and~language identification \cite{zubiaga2016tweetlid}.

NLP techniques specifically adapted for Twitter
have also been used to infer demographic characteristics such as gender,
age or location \cite{cesare2017detection,morgan2017predicting}. Moreover,
relationships, style shifting and community dynamics can also be inferred from
language analysis \cite{nguyen2016computational}. Of particular interest to us
is the~body of work performed with the~objective of age or life stage detection
for Twitter users. Previous works usually generate their own manually annotated
datasets, covering languages such as Dutch, English or Spanish
\cite{rao2010classifying,al2012homophily,nguyen2013old,marquardt2014age,morgan2017predicting,zaghouani2018arap}
for a user range between 300 and 3000. The~best performing systems are those that model life stage
classification as a binary \cite{rao2010classifying,al2012homophily} or ternary
\cite{nguyen2013old,morgan2017predicting} task.

In relation to the study of the social relationships that are generated within
the network, closer to us are those studies that have aimed to identify
communities of users based on their retweets. Among these, one can find studies
about political polarization \cite{conover2011political}, political affiliation
detection \cite{pennacchiotti2011democrats} or even studies about identifying
communities in movements for independence \cite{zubiaga2017stance}.

Finally, there are different research works investigating the use of low resourced
languages within social networks. An investigation
about Welsh (\emph{Cymraeg}) speakers and Twitter, shows
that speakers of this language are also active in social media
\cite{doi:10.1080/01434632.2013.812096}. Additionally, there is another
work that extracts and analyzes more than 80k tweets in Irish
(\emph{Gaeilge}) to do content, sentiment and network analysis
\cite{doi:10.1080/01434632.2018.1450414}. It is also interesting a study
combining Welsh, Irish and Frisian (\textit{Frysk}) to investigate the use of
hashtags across 3,000 different tweets
\cite{doi:10.1080/01434632.2018.1465429}. All these works show the potentiality
of Twitter to provide text data even for low resourced languages, giving the chance to
find and study a huge variety of languages and cultures.

\section{Heldugazte-Age: A New Dataset for Life Stage Classification}\label{sec:heldugazte-age}

In this section we propose a new methodology to semi-automatically obtain
labeled data to develop life stage classifiers. The result is a new dataset for
to train classifiers for Life Stage Detection, namely, the
\emph{Heldugazte-Age} corpus. 

The first step to identify online communities of young Basque speakers is collecting the data. As a starting point we will use the Basque corpus \textit{Heldugazte} \cite{fernandez2019large}, which consists of 6M Basque tweets from 8,000 users, collected in May 2018\footnote{\url{http://ixa2.si.ehu.es/heldugazte-corpus/heldugazte-osoa.tar.gz}}. In this collection, the last 3,200 tweets from each user were retrieved (if available), including personal tweets and retweets.

The \emph{Heldugazte} corpus will be used to semi-automatically generate a labeled
subset of the corpus, 80K tweets, to train classifiers to detect young/adult
users. The obtained classifiers will then be applied to the rest of the
\emph{Heldugazte} corpus to obtain a large number of
tweets written by young users. This data will be used to detect the communities
between young users.

In order to obtain a young/adult classifier we need some labeled data for
training and evaluation. However, labeling users' tweets by life stage is a difficult
task, due to two main reasons: (i) users age hardly ever appear in the tweets
metadata and, (ii) manually annotating tweets by life stage is far from being
trivial. Examples (1-3) below illustrate the difficulty of manually labeling
individual tweets by life stage and without any additional context. 

\begin{itemize}
	\item[(1)] ``Zarauzko triatloian izena ematea lortu gabe, motibazioa falta'' \textit{I have not managed to sign up for the Zarautz triathlon, I am unmotivated}
	\item[(2)] ``A zer nolako eguraldi kaxkarra ez al du gelditu behar edo'' \textit{What a bad weather, shouldn’t stop or what.}
	\item[(3)] ``5 mila euro, bideo kamera eta telefono mugikor bat eroan dituzte
		lapurrek'' \textit{5,000 euros, a video camera and a cell phone were taken away by the burglars.}
\end{itemize}

In order to overcome this problem, previous sociolinguistic work has argued
that writing style could be associated to author's life stage, assuming that
young people's style is more informal than that of adults
\cite{rao2010classifying,al2012homophily,nguyen2013old,morgan2017predicting}.  

Based on these previous works, Fernandez de Landa et al. \shortcite{fernandez2019large} trained various
classifiers to distinguish between formal and informal writing style in tweets.
Every tweet for every user in the \emph{Heldugazte} corpus was automatically
tagged, projecting from formal/informal to young/adult classification depending
on the concentration of formal/informal tweets in each user's timeline. The
problem with this procedure was to objectively define a threshold for the
proportion of formal/informal tweets required to classify a user as young or
adult. The proposed ad-hoc solution, establishing that if 45\% of the tweets were labelled
informal then the timeline was to be classified as young (adult otherwise) was
far from ideal.

\begin{figure}[!ht]
	\centering
\includegraphics[scale=0.4]{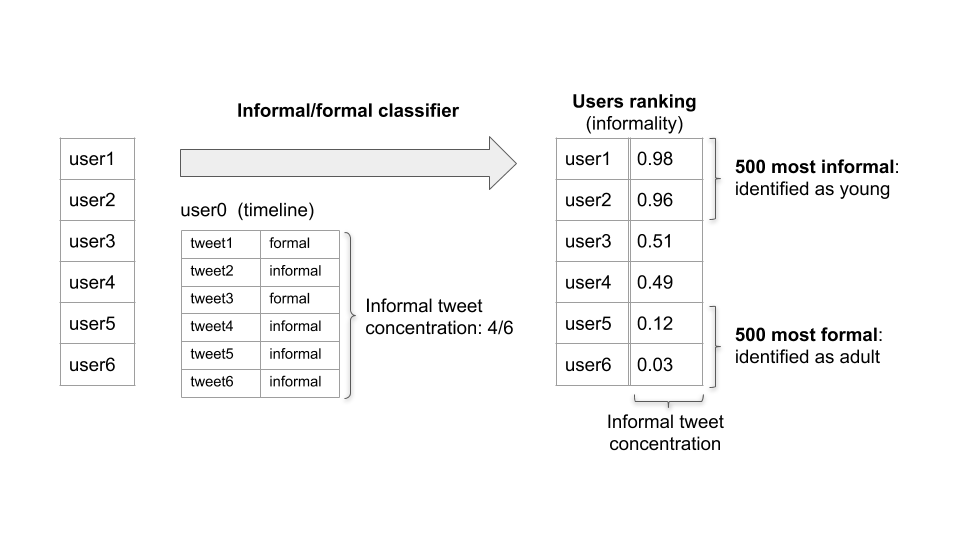} 
\caption{Ranking users by proportion of formal and informal
tweets.}\label{fig:ranking}
\end{figure}

In this paper we propose a new method to objectively and semi-automatically
obtain the labeled data required to train young/adult classifiers. The
procedure is illustrated in Figure \ref{fig:ranking}. 
First, we automatically tagged the 6M tweets in the \emph{Heldugazte} corpus using the
formal/informal classifiers developed by Fernandez de Landa et al.
\shortcite{fernandez2019large}. Second, we ranked users according to the
proportion of informal tweets in their timeline. The top users would contain
mostly informal tweets whereas the users at the bottom of the rank would
consist mostly of formal tweets. Third, a manual inspection of 100 timelines
(50 young and 50 adult) established that
it was feasible to manually annotate users at both ends of the ranking as
young/adult. In this step it was particularly helpful to perform the annotation at user-level
because a full timeline provides more contextual information to characterize a
specific user. Fourth, we took the 500 users at the top of the ranking to be
young users and the 500 at the bottom to be adult. As a result, following this
new method we obtain a set of 1,000 users (out of the original 8K users) classified as adult/young based on the initial
formal/informal manual categorization of 1,000 tweets provided by
Fernandez de Landa et al. \shortcite{fernandez2019large}.

The final step consisted of randomly sampling a number of tweets per user. The
idea was to vary the number of tweets and topics available per user providing 
a sample general enough to train robust young/adult
classifiers. With this objective in mind, we picked 100 random tweets per user
(if a user's timeline did not contain at least 100 tweets then we used the full
timeline) assigned to each of them the label attributed to the user
(young/adult). As it is shown in Table \ref{tab:user-lifestage}, the final labeled set, \emph{Heldugazte-Age}, contains 80K
tweets equally distributed into the young and adult classes. The data was
splitted for experimentation into a training (60\%), development (20\%) and
test (20\%) set, resulting, for each class, in 24K tweets for training and 8K for development and
test, respectively.

\begin{table}[ht]
  \centering
  \begin{tabular}{lrrr} \hline
    & young & adult & total\\\hline
    users & 500 & 500 & 1,000 \\
    tweets & 40,000 & 40,000 & 80,000 \\\hline
  \end{tabular}
  \caption{Annotated corpus for life stage detection at user level.}
  \label{tab:user-lifestage}
\end{table}

\section{Life Stage Classification Systems}\label{sec:experimental-setup} 

Here we present the two main systems used for life stage
detection: (i) an off-the-shelf system based on linear classification and
clustering features \cite{agerri_rigau}, and (ii) a deep learning approach
based on learning contextual, vector-based word representations and the
Transformer architecture \cite{Devlin19}.

Previous approaches address life stage detection as a supervised text
classification task
\cite{rao2010classifying,al2012homophily,nguyen2013old,morgan2017predicting}.
This means that classifiers will learn, from annotated data, that a given tweet
is written by a young or an adult person. An example of the dataset annotations
used for training can be seen in Table \ref{tab:hga-example}. The
\emph{Heldugazte-Age} dataset developed in the previous section will therefore
be used to train three different text classifiers: (i) IXA pipes
\cite{agerri2014ixa} (ii) multilingual BERT \cite{Devlin19} and, (iii) BERTeus
\cite{agerri-etal-2020-give}.

\begin{table}[ht]
  \centering
  \begin{tabular}{rl}
    \hline
    \textbf{Label} & \textbf{Content} (tweet) \\ \hline \hline
    adult & Taldeak mikel laboaren lanean oinarritu du bere hurrengo diskoa. \\
         & \textit{The band has based their next album on the work of Mikel Laboa.} \\ \hline
    adult & Gure herriko ateak zabalik dituzu. \\
             & \textit{The doors of our town are opened.} \\ \hline
    young & Buaa q follaa eun guztia eon zea ikasi ordez jolasateeenn jajaja. \\
             & \textit{How lucky! You have been all day playing instead of studying hahaha.} \\ \hline
    young & Batzutan ze gutxi aguantatze zaituten. \\ 
             & \textit{Sometimes I can't stand you.} \\\hline
  \end{tabular}
  \caption{Examples taken from the \emph{Heldugazte-Age} dataset.}
  \label{tab:hga-example}
\end{table}

\subsection{IXA pipes}\label{sec:ixapipes}

IXA pipes is a set of tools with a multilingual approach across NLP
tasks. This system has been successfully used in several
sequence labelling tasks for various languages, including Named Entity
Recognition \cite{agerri_rigau}, and Opinion Target Extraction
\cite{agerri2019language}.

The general objective of IXA pipes is to provide a general semi-supervised approach that performs well across languages and tasks. This approach consists of two different components. In the first one a set of linguistically shallow features are extracted from the local context; these features are based on orthographic and ngrams and character-based information to capture multi-word patterns and prefixes and suffixes of words, which has proven useful to work with an agglutinative language such as Basque \cite{agerri_rigau}. The second, semi-supervised, component injects external knowledge previously obtained from the unsupervised induction of clustering models over large amounts of texts. This component provides several benefits. First, it generates denser document representations, given that a document is represented with respect to the number of dimensions (clusters) specified in the obtained clustering model. Second, by training the clustering models on source data from different domains and text genres it is possible to inject domain-specific knowledge into the system. Finally, IXA pipes includes the possibility of including features from three types of clustering models \cite{brown1992class,clark2003combining,mikolov2013distributed}, which helps to represent domain-specific information via complementary semantically induced knowledge. More details can be found in \cite{agerri_rigau,agerri2019language}. For this particular work
we train the IXA pipes document classifier using the same experimental setup used in
\cite{fernandez2019large}.

\subsection{Transformer Models}\label{sec:transformers}

As for many other NLP tasks, current best performing
systems for text classification are large pre-trained language models which allow to build rich
representations of text based on contextual word embeddings. Deep learning methods in NLP represent words as
continuous vectors on a low dimensional space, called word embeddings. The
first approaches generated static word embeddings
\cite{mikolov2013distributed,fasttext1_bojanowski2017enriching}, namely, they
provided a unique vector-based representation for a given word independently of
the context in which the word occurs. This means that polysemy cannot be
represented. Thus, if we consider the word `bank', static word embedding
approaches will generate only one vector representation even though such word
may have different senses, namely, `financial institution',`bench', etc. 

In order to address this problem, contextual word embeddings were proposed. The
idea is to be able to generate word representations according to the
context in which the word occurs. Currently there are many approaches to
generate such contextual word representations, but we will focus on those that
have had a direct impact in text classification, namely, the models based on the Transformer architecture \cite{vaswani2017attention}
and of which BERT is perhaps the most popular example \cite{Devlin19}.

There are several multilingual versions of these models. Thus, the multilingual
version of BERT \cite{Devlin19} was trained for 104 languages. More recently,
XLM-RoBERTa \cite{conneau2019unsupervised} distributes a multilingual model
which contains 100 languages. Both
include Basque among the languages.

These multilingual models perform very well in tasks involving
high-resourced languages such as English or Spanish, but their performance
drops when applied to low-resourced languages \cite{agerri-etal-2020-give}.
Although this is still an open issue, a number of reasons can be found in the
literature. First, each language has to share the quota of substrings and
parameters with the rest of the languages represented in the pre-trained
multilingual model. As the quota of substrings partially depends on corpus
size, this means that larger languages such as English or Spanish are better
represented than lower resourced languages such as Basque. Moreover, multilingual models also seem to behave
better for structurally similar languages \cite{karthikeyan2020cross}.

BERTeus \cite{agerri-etal-2020-give} is a language model trained in Basque language following
BERT's architecture \cite{Devlin19}. They show that training a
monolingual Basque BERT model obtains much better results than the
multilingual versions. In this paper we will compare the performance of multilingual BERT and BERTeus
for life stage detection using the same hyperparameters as in Agerri et al.
\shortcite{agerri-etal-2020-give}.

\section{Life Stage Detection}\label{sec:life-stage-detection}

In this section we will use the \emph{Heldugazte-Age}
corpus to train the classifiers previously described. The best
classifier will then be applied to the whole \emph{Heldugazte} dataset in order
to obtain a young/adult classification of the 8K Basque tweet users contained
in the corpus. Additionally, an analysis of the results is performed to better
understand the quality of the semi-automatically obtained annotations.

\subsection{Experimental Results} 

It should be noted that, in contrast to our previous work
\cite{fernandez2019large}, the
\emph{Heldugazte-Age} corpus allows us to directly classify users as young/adult,
without having to perform the formal/informal step. 

We perform minimal pre-processing on the tweets; we remove URLs, hashtags and
usernames, leaving label-tweet pairs such as the examples shown in
Table \ref{tab:hga-example}. This procedure has proven to be useful in previous
text classification works with tweets \cite{agerri-etal-2020-give,zotovasemi}.

Table \ref{tab:bert-garapen}  
reports the results obtained using the three systems described in Section
\ref{sec:experimental-setup}. The high scores show that our semi-automatic method to obtain young/adult
training data produces good quality annotations. Furthermore, the differences between the
systems are not that large, although BERTeus is consistently the best scoring
model. 

\begin{table}[ht]
  \centering
  \begin{tabular}{lcccc}
    \hline
    System & Accuracy & Precision & Recall & F1 Score\\ \hline \hline
    IXA pipes & 0.956 & 0.977 & 0.935 & 0.955 \\
	mBERT & 0.955 & 0.972 & 0.936 & 0.954 \\
    BERTeus & \textbf{0.963} & 0.968 & 0.958 & \textbf{0.963} \\ \hline
  \end{tabular}
  \caption{Evaluation results of young/adult classifier models on the
  \emph{Heldugazte-Age} test set.}
  \label{tab:bert-garapen}
\end{table}

In order to further test the robustness of our semi-automatic method, described
in Section \ref{sec:heldugazte-age}, we decided to manually
annotate 200 randomly selected tweets. Two human annotators labeled the 200
tweets and we calculated an agreement between the annotators of 0.78 and a Kappa
score of 0.55, showing a moderate agreement between them. Furthermore, the
accuracy of the two annotators are 0.795 and 0.775 respectively. When
comparing these scores with the results reported in Table \ref{tab:bert-garapen},
it is clear that manually annotating young/adult at tweet level is a very
difficult task. These results also show the effectiveness of our method to
obtain the \emph{Heldugazte-Age} corpus.

In the rest of this paper we will use the BERTeus fine-tuned model to automatically annotate
the whole \emph{Heldugazte} corpus. It should be noted that the classifier
works at tweet level (as shown by Table \ref{tab:hga-example}). This means that
once every tweet is automatically annotated, we still need to decide whether each of
the user timelines corresponds to a young or adult user based on the number of
individual tweets classified as young/adult.

\subsection{Labeling the Large Corpus} 

Once the tweet classifier is ready to use, we apply the following strategy to automatically annotate the tweets in the \emph{Heldugazte} corpus. 
First we assign a discrete \emph{young} or \emph{adult}
label to each tweet. We then obtain a single score by averaging the number of the young/adult
classified tweets of each user's timeline. 

The last step is to decide whether a given timeline corresponds to a young or
adult user based on the score obtained from the classification of the
individual tweets. In order to avoid establishing an ad-hoc value as a
threshold, we introduce a third class for classification. In other words, a new
synthetic category, ``underdetermined'', is created thus transforming a binary
task into a ternary one. 

Based on the new ternary task, two thresholds are used instead of one, located at
60\% and 40\% of the number of tweets annotated as young in each
timeline. Thus, if the proportion of labels or the average
probability is over 60\%, the user will be defined as a young user. On the
other hand, if those values are lower than 40\%, the timeline will be
considered to be from an adult user. Finally, if the value is between 40\%
and 60\% we will consider the timeline to be ``underdetermined'', meaning that we
do not have enough evidence to decide the life stage of the user. Adding the
\emph{underdetermined} class has the benefit of avoiding to commit ourselves
to classify difficult cases as young/adult. 

We are also interested in comparing the distribution of young/adult users obtained
using the described procedure with those that are obtained using our previous
method \cite{fernandez2019large}. As a reminder, in our previous work each
tweet is classified as formal/informal and then, based on the number of
informal tweets we decide whether the user is young or adult. However,
for a fair comparison we will adapt it to use two thresholds (60/40 for
young/adult) and three classes, as it has been described above. 

\begin{table}[ht]
  \centering
  \begin{tabular}{lccc} \hline
     & Adult & Underdetermined & Young \\ \hline \hline
    Previous system & 5,213 & 911 & 962 \\
    New system & 4,472 & 980 & 1,635 \\ \hline
  \end{tabular}
  \caption{Classifying users in terms of age stage (young/adult).}
  \label{tab:userclassi}
\end{table}

Table \ref{tab:userclassi} shows the number of timelines classified as
young/adult or underdetermined using our new and old methods. It can be seen
that the main difference corresponds to the quantity of young users obtained by each of the
methods. In the next subsection we further look into this issue.

\subsubsection{Comparison of Methods} 

\begin{figure}[ht]
\begin{floatrow}
\ffigbox
    {\includegraphics[width=\linewidth]{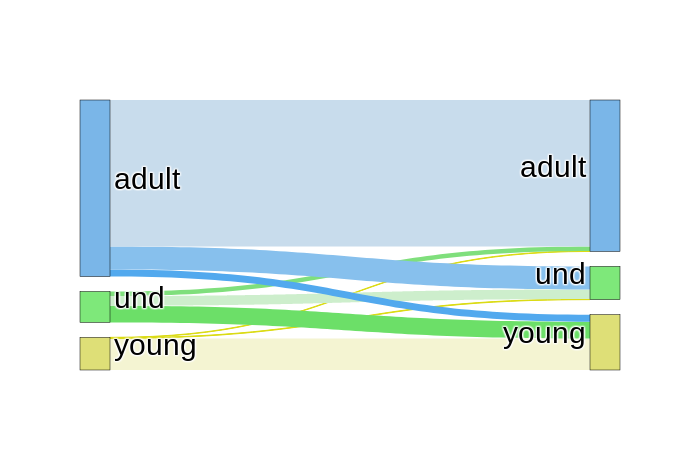}}
    {\caption{Previous system to new system.}
    \label{fig:oldtonew}}
\capbtabbox{%
  \begin{tabular}{lc} \hline
     & previous to new \\ \hline \hline
    adult to adult & 4325 \\
    adult to und* & 679* \\
    adult to young* & 209* \\\hline
    
    und to adult & 133 \\
    und to und & 285 \\
    und to young* & 493* \\\hline
    
    young to adult & 14 \\
    young to und & 16 \\
    young to young & 933 \\ \hline\hline
  \end{tabular}
  }
  {\caption{Variations between previous and new systems classifications.}
  \label{tab:userlabelingdif}}
\end{floatrow}
\end{figure}

In this section we look at those variations in the automatic annotations
assigned by the old method \cite{fernandez2019large} with respect to the one
presented in this paper. Table \ref{tab:userlabelingdif} shows the differences
of classifying the timelines using the old (based on formal/informal
classification of tweets) with respect to the the new method (based on
young/adult classification). After a superficial look to the variations, 
it can be seen that 21.79\% of the labels were differently labeled from
previous to new system, a substantial difference. Besides, one of the most
significant variations is the increase in the amount of users labeled as young.

Two of the three most important variations, marked with an asterisk, refer to the transfer
of timelines to the young class. It is also important the transfer from
\emph{adult} to \emph{underdetermined}. Taking a deeper look into these specific
cases, we manually inspected some randomly chosen timelines to see if these
transfers are actually true positives or whether they are misclassifications.
The objective of this comparison was to study the transfer of classifications
across categories (from adult to young, for example) when using the new
classification method. More specifically, we analyzed a random sample of 10\% of
the cases for each variation. 

Below we can see example tweets from three different users. With respect to
@user2 and @user3, they show that our new method, as opposed to the old one,
actually classifies correctly their timelines. Thus, by looking at their tweets
it seems that the users are indeed young, based on the writing style but also
because the tweets talk about exams, an activity usually related to young
people. The case of @user1 is more contentious, as it seems too difficult to
establish the life stage of the user based on those examples, which is why the
\emph{underdetermined} classification does not seem misplaced.

\begin{itemize}
	\item  Adult to underdetermined variation example, \textit{@user1}:
    \begin{itemize}
    \item[--] \textit{tweet\_1a}: A zer nolako eguraldi kaxkarra ez al du gelditu behar edo. \textit{What a bad weather, shouldn't stop or what.}
    \item[--] \textit{tweet\_1b}: Gu erakusteko prest, etorri daitezela lasai eskuzabalik hartuko ditugu eta. \textit{We are ready to show it, we will wait for them with open arms.}
    \end{itemize}

\item  Adult to young variation example, \textit{@user2}:
    \begin{itemize}
    \item[--] \textit{tweet\_2a}: Horrelakoekin gustua ta guzti hartzen zaio ikasteari. \textit{With this, you take pleasure in learning.}
    \item[--] \textit{tweet\_2b}: Buenobueno ba ikasiko dut gehio jaja ta ikusikozu gaindituko dutt jaja. \textit{Weeeell weeeell, I'll learn more haha and you'll see if I can pass the exam haha.}
    \end{itemize}

\item  Underdetermined to young variation example, \textit{@user3}:
    \begin{itemize}
    \item[--] \textit{tweet\_3a}: Ze txupi txatxi no me da la nota. \textit{Awesome I don't get to pass...}
    \item[--] \textit{tweet\_3b}: Ai naiz rayatzen pixkat asko con la mierda de la uni. \textit{Oh I'm going crazy a little bit with university shit.}
    \end{itemize}
\end{itemize}

Figure \ref{fig:oldtonew} depicts the variations in the classification from the
using the old method (left) with respect to the new one (right). The manual
inspection performed would indicate that such variations are in fact
correct. In other words, it would seem that the method introduced in Sections
\ref{sec:experimental-setup} and 4.2 to develop classifiers to automatically annotate users in the \emph{Heldugazte} 
corpus as young/adult/underdetermined produce better results.

\section{Relationship network}\label{sec:relationship-network}

In this section we will study the relations that appear between Basque young
Twitter users.  The starting point will be the retweets of messages written in
Basque by the 1,635 users classified as \emph{young} in the previous section.
We select the retweets because they are the type of interactions between users
that can show correlations better than other interactions such as mentions
\cite{conover2011political}. Specifically, from the 418,903 retweets of the
1,635 young users, we extracted 24,837 nodes and 148,304 edges or connections.
The nodes correspond to the users doing the retweets (our sample of 1,635
users) but also different users receiving them (from our sample or not). On the
other hand, the edges represent if a source user has retweeted one or more
times another target user, representing the connections in the graph.

Once the retweets are gathered we proceed to transform the unstructured data
into a readable graph. First, we created a giant graph using the data
(retweets) from each user. To build the graph, two features extracted from each
retweet were used: (i) the retweeter and (ii) the user retweeted. After extracting
the data, the visualization of the graph was created using the
\textit{gephi} program \cite{bastian2009gephi}. Second, we gave the network a
spatial structure by using the \textit{ForceAtlas2} algorithm
\cite{jacomy2014forceatlas2}, ordering the nodes according to the established
relations. This algorithm displays a spatialization process, giving a readable shape to a network with the aim of transforming the network into a map. This technique simulates a physical system in order to spatialize a network. As a result of this process, those nodes that are unrelated repulse each other, while related ones will attract each other. The algorithm can turn structural proximities into visual proximities, allowing the analysis of this particular type of data based on interactions. Thus, the relations can be displayed in a (huge) graph.

After creating the graph, we focused on two different aspects.  First we
identified the most important nodes of the network, to establish which users
are the most influential. In a second step we uncovered the implicit
communities of Basque users, splitting the huge graph into more readable
subgroups that allowed us to infer the communities of young people.

\subsection{Basque influencers among young users}\label{sec:relat-nodes}

The most retweeted users of the graph can reveal important characteristics of
the investigated sample. The most important nodes show which users are the
leaders for our sample. Thus, in Table \ref{tab:taldeak} we can see the
top 15 most retweeted users, based on two different
classifications. On the one hand, there are those users with most overall retweets
(Table \ref{tab:taldeaka}). On the other hand, we have the users that have been
retweeted by different young users, focusing on how many different users have
retweeted these users (Table \ref{tab:taldeakb}). These two rankings illustrate
which users are actually the most influential between young Basque Twitter
users.

\begin{table}[h!]
           \centering
           \captionsetup[subtable]{position = below}
          \captionsetup[table]{position=top}
           \begin{subtable}{0.45\linewidth}
               \centering
               \begin{tabular}{lcl}\hline
                   User & Times retweeted \\ \hline \hline
                   @berria & 8671 \\
                   @argia & 5646 \\
                   @ernaigazte & 4553 \\
                   @topatu\_eus & 4236 \\
                   @enekogara & 3274 \\
                   @naiz\_info & 3262 \\
                   @ZuriHidalgo & 2568 \\
                   @AskeGunea & 2561 \\
                   @RealSociedadEUS & 2531 \\
                   @larbelaitz & 2471 \\
                   @ArnaldoOtegi & 2188 \\
                   @iBROKI & 2031 \\
                   @LeakoHitza & 1893\\ 
                   @athletic\_eus & 1818\\
                   @euskaltelebista & 1744\\ \hline
               \end{tabular}
               \caption{Total RTs done to users.}
               \label{tab:taldeaka}
           \end{subtable}
           \hspace*{0.1cm}
           \begin{subtable}{0.45\linewidth}
               \centering
               \begin{tabular}{lcl}\hline
                   User & Users retweeted\\ \hline \hline
                   @berria & 998 \\
                   @argia & 844 \\
                   @naiz\_info & 710 \\
                   @larbelaitz & 585 \\
                   @topatu\_eus & 531 \\
                   @ArnaldoOtegi & 518 \\
                   @ernaigazte & 478 \\
                   @enekogara & 454 \\
                   @HamaikaTb & 442 \\
                   @jpermach & 427 \\ 
                   @axierL & 413 \\
                   @ielortza & 407 \\
                   @MaddalenIriarte & 404 \\
                   @boligorria & 398 \\
                   @GureEskuDago & 394 \\ \hline
               \end{tabular}
               \caption{Young users retweeted accounts.}
               \label{tab:taldeakb}
           \end{subtable}
           \caption{Most retweeted accounts by young users.}
           \label{tab:taldeak}
       \end{table}

By looking at the obtained rankings, we can see that at the top there are
accounts related to Basque media: @berria (newspaper), @argia (weekly magazine), @naiz\_info (newspaper), @topatu\_eus (digital media related to young people), @HamaikaTb (a television channel),
@euskaltelebista (Basque public television broadcaster) and @LeakoHitza (local newspaper); and Basque journalists: @larbelaitz, @axierL, @boligorria (the three of them journalists from Argia) and @iBROKI (sports journalist in the Basque Television). We attribute this to the fact that those people perform important roles in the
creation and distribution of Basque language content in the Web. 

After a manual analysis of the obtained influencers, it has to be said that only two of them are accounts related to young people: @ernaigazte  and @topatu\_eus are both accounts related to organizations formed by young people. On the one hand, @ernaigazte is the account of the Basque nationalist left youth organization, named Ernai. On the other hand, @topatu\_eus is a digital media account related to young people, very related to the Basque nationalist left. The lack of influencers among young users, could be related to the characteristics of Twitter, which is mostly structured around to political issues.

\subsection{Basque speaking communities for young users}\label{sec:relat-young-subgr}

\begin{figure}[ht]
  \centering
  \begin{subfigure}[b]{0.48\linewidth}
    \includegraphics[width=\linewidth]{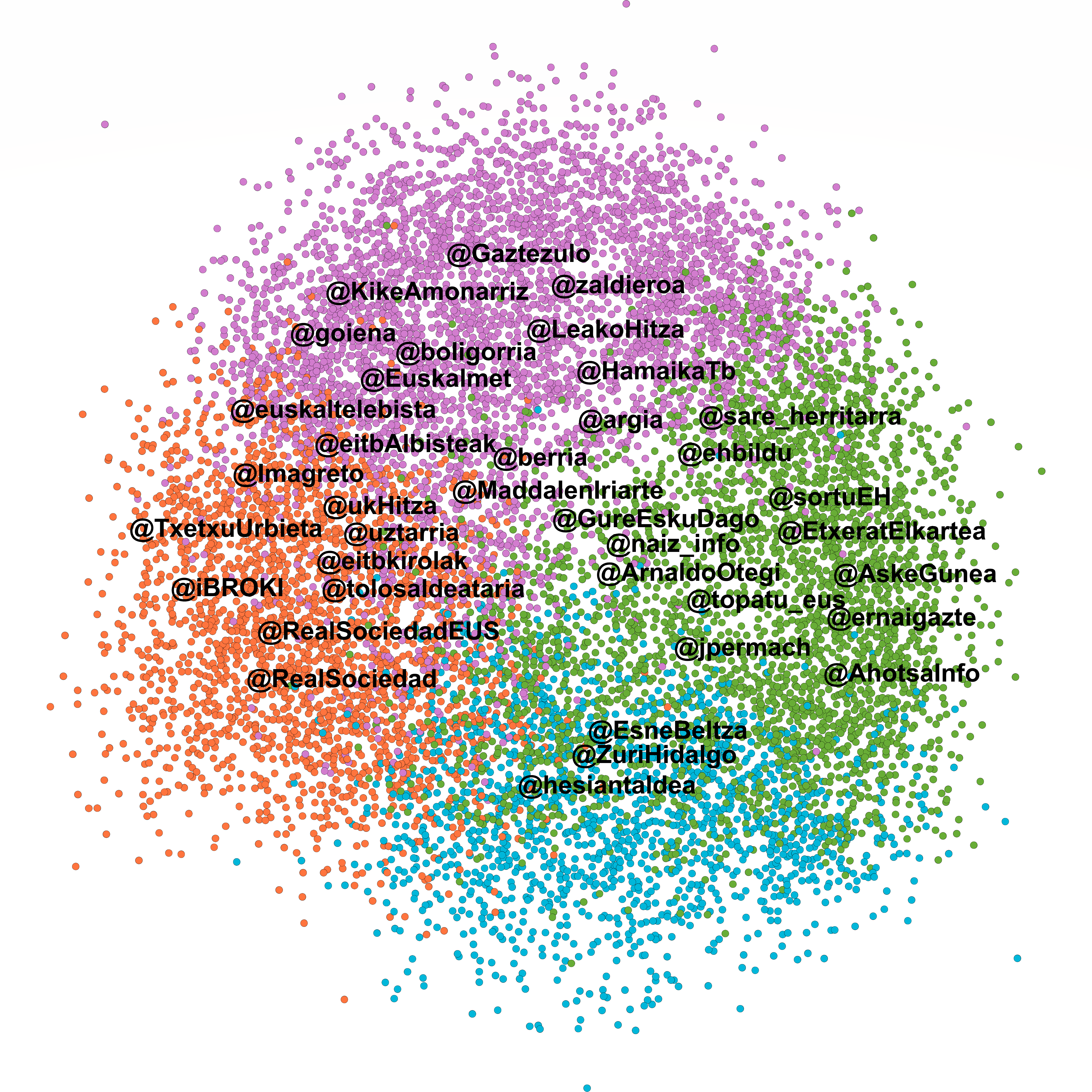}
    \caption{Young users communities, some important nodes.}
    \label{fig:young relations nodes}
  \end{subfigure}
  \hspace*{0.1cm}
  \begin{subfigure}[b]{0.48\linewidth}
    \includegraphics[width=\linewidth]{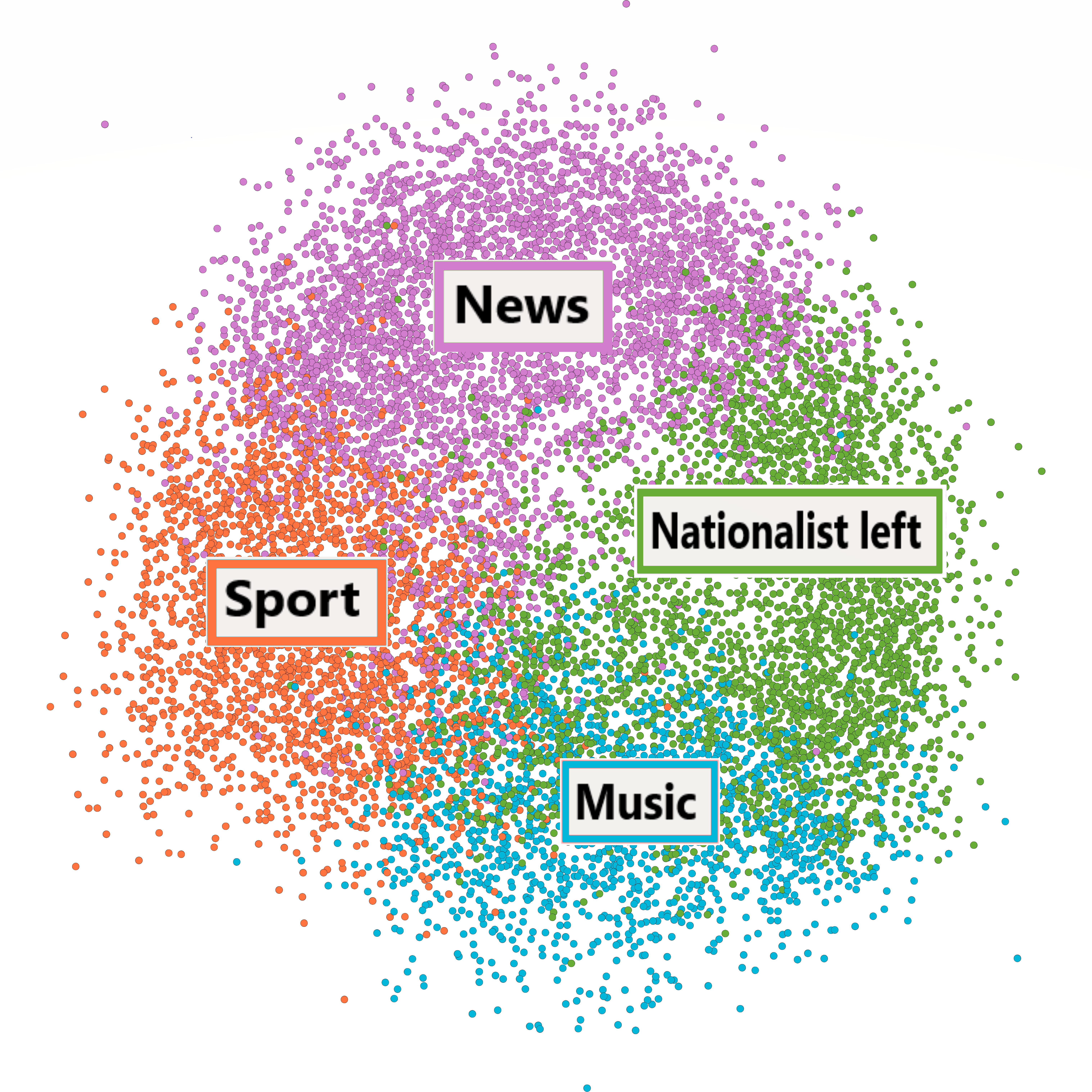}
    \caption{Young users communities, topic.}
    \label{fig:young relations topics}
  \end{subfigure}
  \caption{Young users graph divided in communities.}
  \label{fig:young relations}
\end{figure}

Once the main network graph was created, we split it into subgroups to
analyze how the sub-communities or subgroups inside each one are formed. We divided each
graph into subgroups using the node2vec algorithm \cite{grover2016node2vec},
which allows us to obtain consistent subgroups. The node2vec algorithm can
freely explore network neighborhoods which is useful to discover homophilic communities. Unlike modularity based algorithms \cite{blondel2008fast}, used in a previous analysis of Basque communities \cite{fernandez2019large}, node2vec gives the opportunity to choose the exact number of communities to be extracted. Besides, this algorithm can be tuned in order to give more importance to homophilia or to structural equivalence. Thus, Figure \ref{fig:young relations} shows that node2vec generates clearly distinguishable sub-communities, which in turn makes them more interpretable thereby facilitating the understanding of the existing relations between them.

After splitting the graph into four communities, we had to infer the
main characteristics of each subgroup. For this process we focus again on the
most important nodes which are the ones used to define the community itself.

Each of the subgroups displays a common characteristic, namely, all of them
have a direct relation with topics or issues related to the Basque Country.
Those topics are different in each of the subgroups in the graph, showing the
characteristics or differences of each community.  Thus, it can be seen that
Basque language interactions are used to talk about various Basque current
affairs (news) and politics (Nationalist left). Also, it can be seen that music
and sports are also widely commented by young people. In other words, it seems
that the main function of Twitter interaction is to spread content about
politics and social issues but with a clear focus on the Basque community and
language. In the following we describe the main characteristics of each of the
four subgroups contained in the graph. 

\begin{itemize}

\item \textbf{News} (29.96\%): In this community, the nodes found at the top of the ranking
	are related to news media from the Basque Country (@berria, @argia, @HamaikaTb,
	@eitbAlbisteak, @euskaltelebista, @zuzeu, @euskadi\_irratia, @Gaztezulo,
	@Sustatu, @eitbeus...), specific Basque journalists (@MaddalenIriarte,
	@boligorria, @urtziurkizu, @zaldieroa, @bzarrabeitia, @AneIrazabal... ) and
	also to other very active users that write about the most noteworthy news in Basque (@ielortza,
	@kalaportu, @KikeAmonarriz, @maia\_jon...). 

\item \textbf{Nationalist left} (26.98\%): The composition of this particular subgroup is characterized by nodes related, in different ways, with the nationalist/independentist Basque left. The nodes can refer to news media (@naiz\_info,
	@topatu\_eus, @info7irratia, @AhotsaInfo...), political and social organizations (@ernaigazte,
	@GureEskuDago, @AskeGunea, @ehbildu, @sortuEH, @EtxeratElkartea...) and
	politicians from the main parties in this political movement (@ArnaldoOtegi, @jpermach...).

\item \textbf{Sports} (22.58\%): In the Sports subgroup the most important nodes are actually journalists (@iBROKI, @XabierEuzkitze,
	@Imagreto, @TxetxuUrbieta, @jontolest, @unaizubeldia...) and news media
	(@eitbkirolak, @ukHitza, @3ErregeenMahaia...) specifically specialized in the sports domain. Thus, for this specific group the top accounts also refer to newspapers and television broadcasters. Other important nodes here are those related to sport teams, such as football teams (@RealSociedad, @RealSociedadEUS, @SDEibar,
	@AthleticClub...) and Basque ball clubs (@ASPEpelota...) or their players, which are mostly footballers from professional teams (@InigoMartinez, @mikelsanjo6, @ilarra4...) or even well known cyclists (@AmetsTxurruka, @mikelastarloza, @Markelirizar...). 

\item \textbf{Music} (20.49\%): In the Music subgroup appear in prominent places music bands or singers which sing in Basque (@ZuriHidalgo,
	@vendettaska, @hesiantaldea, EsneBeltza, @gatibu, @ZeEsatek...), although other accounts related to music seem to be also very active (@GustokoMusika, @euskalkantak5,
	@KantuBatGara...).

\end{itemize}

Figures \ref{fig:young relations nodes} and \ref{fig:young relations topics}
show that young Basque users generally interact with users related
to social issues (politics and news) as well as with those related to leisure
(music and sport). Due to the new method applied for community detection, we are able to map the communities in a more consistent way, showing in a clear way where each community is located. The position of each community on the graph and the
closeness between communities show how related the topics are between them.
In this way we can see that communities related to social issues are next to
each other, while the same occurs with the leisure-related communities. The community related to politics is
close to news and music, illustrating both the relation with current news
and the political stance of some Basque music bands. 
Besides, in three of the four communities (News, Nationalist left and Sports), media and journalists are referential, proving again that media is important at disseminating Basque content among young people, in spite of the main topic of the community.

In this section we show that combining the community
detection algorithm and the visualization of the spacial representation of the
graph, humans can easily interpret the meaning and characteristics of the
displayed data. Thus, any information based on user interactions could be displayed
and interpreted using these techniques, helping us to transform unstructured
information into knowledge.

\section{Concluding Remarks}\label{sec:conclusion}

In this paper we have presented a new methodology to perform demographic analysis by
processing a large amount of tweets in Basque language. We have applied machine
learning and deep learning approaches to Natural Language Processing to extract
structured knowledge from unstructured data. 

Our experimental results have shown that our new method produces good quality
labeled data for training young/adult classifiers. This allows us to generates
a new dataset of 80K tweets annotated at user level, namely,
\emph{Heldugazte-Age}. The analysis of the
classifiers performance has shown that, when compared with manual annotations at
tweet level, the annotations of our semi-automatically generated \emph{Heldugazte-Age} dataset
benefit from taking into account user-level information. Furthermore, we
have experimented with modern deep learning techniques for NLP and for the
detection and visualization of communities in Twitter. The use of these
technologies has allowed us to get more consistent
and readable results than in our previous approach \cite{fernandez2019large}, apart from a better
understanding of communities and their interactions. 

As a result of our new methodology, we have seen that the the young
Basque users can be grouped in four main communities. Furthermore, we have also
seen that the most influential accounts among young users are related with
Basque media, revealing the importance of this actor at disseminating content in Basque among the youngest. 
A general conclusion has been that Basque
is mostly used in Twitter to speak about Basque-related topics, being that
news, politics, sport or music. 

We believe that the methodology presented in this paper might be of interest for
other NLP tasks and other types of social and demographic studies. Finally, we
publicly distribute every resource (software and data) to facilitate further
research for low resourced languages such as
Basque\footnote{\url{https://github.com/ixa-ehu/heldugazte-corpus}}.

\section*{Acknowledgements}
This work has been partially funded by the~Spanish Ministry of Science, Innovation and Universities (DeepReading RTI2018-096846-B-C21, MCIU/AEI/FEDER, UE), \textit{Ayudas Fundación BBVA a Equipos de Investigación Científica 2018} (BigKnowledge), and DeepText (KK-2020/00088), funded by the Basque Government. Rodrigo Agerri is also funded by the RYC-2017-23647 fellowship and acknowledges the donation of a Titan V GPU by the NVIDIA Corporation.

\bibliographystyle{coling}
\bibliography{coling2020}

\end{document}